\documentclass[epj]{webofc}
\usepackage[varg]{txfonts}   
\usepackage{amsmath,amsfonts,amssymb,bbm}
\usepackage{pdfsync}
\usepackage{showlabels}
\usepackage{graphicx}
\usepackage{hyperref}
\usepackage{mathrsfs}
\newcommand{\bit}{\begin{itemize}}
\newcommand{\eit}{\end{itemize}}
\newcommand{\bd}{\begin{description}}
\newcommand{\ed}{\end{description}}

\newcommand{\bc}{\begin{center}}
\newcommand{\ec}{\end{center}}
\newcommand{\Ref}[1]{(\ref{#1})}

\newcommand{\C}{{\mathbb C}}

\newcommand{\R}{{\mathbb R}}

\def\T{\mathbbm T}

\newcommand{\SU}{\mathrm{SU}}

\newcommand{\SL}{\mathrm{SL}}

\newcommand{\su}{{\mathfrak{su}}}

\renewcommand{\sl}{{\mathfrak{sl}}}

\newcommand{\be}{\begin{equation}}
\newcommand{\ee}{\end{equation}}
\newcommand{\bea}{\begin{eqnarray}}
\newcommand{\eea}{\end{eqnarray}}
\newcommand{\bs}{\begin{subequations}}
\newcommand{\es}{\end{subequations}}

\newcommand{\w}{\wedge}

\newcommand{\tr}{{\rm Tr}}
\newcommand{\f}{\frac}
\newcommand{\tl}{\tilde}

\newcommand{\Id}{\mathbbm{1}}

\newcommand{\re}{\mathrm{Re}}
\newcommand{\im}{\mathrm{Im}}

\renewcommand{\a}{\alpha}  \newcommand{\g}{\gamma}
\renewcommand{\d}{\delta}  \newcommand{\eps}{\epsilon}  
 \renewcommand{\th}{\theta}      
\let\m=\mu  \let\n=\nu  \let\r=\rho \newcommand{\s}{\sigma}       \let\om=\omega
\let\G=\Gamma      

\newcommand{\po}{\pi\om}
\newcommand{\tpo}{\tl\pi\tl\om}

\woctitle{International Conference on New Frontiers in Physics 2013}
\begin{document}
\title{Loop quantum gravity, twistors, and some perspectives on the problem of time}

\author{Simone Speziale\inst{1}\fnsep\thanks{\email{simone.speziale@cpt.univ-mrs.fr}}}

\institute{Centre de Physique Th\'{e}orique, CNRS-UMR 7332, Aix-Marseille \& Toulon Universities, Luminy Case 907, 13288 Marseille, France}

\abstract{%
I give a brief introduction to the relation between loop quantum gravity and twistor theory, and comment on some perspectives on the problem of time.
}
\maketitle
%
\section{Introduction}
\label{intro}

Almost a century after its birth, research in quantum gravity is still a fertile ground for ideas and techniques for theoretical and mathematical physics, with a continuous development of tools that turn out to have broader applications than initially intended. 
Twistor theory is no exception: introduced by Penrose in the early sixties as a mathematical tool to capture the causal structure of spacetime,  it has since found applications way beyond its original purpose, from quantum field theory and analytic S-matrix theory (for a review, see e.g. \cite{AdamoMasonSkinner11}) to non-commutative geometry \cite{SmolinTwistorRL13}. In gravity, twistors have paved the way for studies on the role of self-duality and holomorphicity, out of which came the discovery by Ashtekar that the kinematical degrees of freedom of gravity can be equally carried by a gauge connection, instead of the metric. Gravity is thus a gauge theory in its own proper sense, the equivalence principle being the physical description of the gauge freedom. 
Ashtekar variables are the basic ingredient of loop quantum gravity (LQG), and following this Ariadne's thread, it has recently been found that twistors can be used to describe the phase space of loop quantum gravity. This introduces new tools for loop quantum gravity, and on the other hand, a new way in which twistors can be seen as non-linear gravitons, by relating them to the quanta of space appearing in LQG. 
In this lecture I provide a brief introduction to this topic, and to the perspectives it opens to the problem of time. 

\section{Ashtekar-Barbero variables and quanta of space}
\label{sec-1}
What makes general relativity a field theory different from all others is the fact of being a totally constrained system: 
the canonical analysis, originally performed by Arnowitt, Deser and Misner (ADM) \cite{ADM}, identifies canonically conjugated variables as the 3-metric of a given space-like hypersurface $\Sigma$,
and its extrinsic curvature. These basic variables are however not observable: On this kinematical phase space, one finds 4 first class constraints, generating spacetime diffeomorphisms in the canonical language, thus there are only 2 degrees of freedom in the system.
The presence of constraints in the phase space is common to the description of the other gauge interactions. What is peculiar to general relativity is that on the reduced phase space, the Hamiltonian vanishes, and all observables commute with it.
Therefore, there is no physical evolution associated with the coordinate $t$ used in the canonical analysis. The lack of absolute notions of time and energy encode Einstein's far-reaching diffeomorphism invariance.

This problem of time has two sides to it. The first side, is the purely technical problem of identifying the physical degrees of freedom at the non-perturbative level: this requires finding the general solution to the constraints, which form a system of non-linear PDEs. 
The second side is the heart of the problem, and carries conceptual as well as technical difficulties: due to the vanishing of the Hamiltonian, the constraint-free data are `frozen', and do not evolve with respect to the time parameter. All physics is encoded on the initial surface, including information on the proper time and trajectories of all possible observables, and thus the dynamical descriptions associated with them. Unravelling a more traditional dynamics from this structure is the main challenge of the problem of time. To come to grasp of the conceptual difficulty of the problem, a clarifying approach to the problem is given by Rovelli's partial observables \cite{CarloBook}, where one partially gives up the exclusive use of Dirac observables, and instead seeks to identify suitable clocks, and then describe the evolution in relative terms.
The situation trivialises in many circumstances of physical interest, such as weak fields with asymptotic flatness, or mini-superspace models for cosmology. In these cases, a preferred time is singled out by the working hypothesis, and a meaningful physical evolution immediately obtained.
The holy grail of the problem of time, the one mathematical relativists are after, is to find a solution to this question in general terms.

A second, more formal, sense in which general relativity differs from the other descriptions of forces is in the use of the metric as fundamental field, that is a tensor instead of a gauge connection. 
The theory does carry a notion of gauge connection, which is the Levi-Civita connection $\G[g]$ used to define parallel transport on a Riemannian manifold. This quantity can be used to bridge between the two descriptions of forces, an idea that dates back to Einstein himself, and to the first attempts to a unification of gravity and electromagnetism by Kaluza and Weyl.
In this logic, one views the Riemann tensor as the curvature tensor of an arbitrary affine connection $\Gamma$. This can be done harmlessly: thanks to an identity first due to Palatini, and to the symmetries of the Riemann tensor, the independent variation of the action with respect to $\G$ and to the metric gives back the Einstein equations. 
With an arbitrary $\G$, one can also consider a second dimension-2 invariant in the action, the completely antisymmetric contraction $\eps^{\m\n\r\s}R_{\m\n\r\s}[\G]$. This gives no contribution to the field equations as it vanishes identically when $\G=\G[g]$. 
Being consistent with all symmetries and of the same dimensions of the Ricci scalar, it should be added to the Einstein-Hilbert action, with its own coupling constant.
In units $G=1$, its coupling constant coincides with the Immirzi parameter of LQG,\footnote{Up to a possible extra contribution coming from the topological Pontryagin term.} and we'll thus refer to it as $\g$. 

A natural way\footnote{Indeed, necessary, if one is to couple fermions to gravity.} to bring the gravitational connection to the forefront is to use Cartan's tetrad $e^I_\m$, defined via $g_{\m\n}=e^I_\m e^J_\n \eta_{IJ}$, and a connection on a Lorentz bundle, $\om^{IJ}_\m$, taken as an independent variable. The action for general relativity then reads
\be\label{S}
S[e,\om] = \int \tr \big[(\star +\f1\g \Id) \, e\w e\w F(\om)\big],
\ee
where the trace is over internal indices, and $\star=1/2\eps^{IJ}_{KL}$ is the internal Hodge star.
The first term is the rewriting of the Einstein-Hilbert term in terms of tetrads, and it is usually referred to as Einstein-Cartan, or Einstein-Cartan-Kibble-Sciama action. The second term corresponds to the second invariant mentioned above, and it is often referred to as Holst term in the literature. While this term plays no role classically, it is a result of loop quantum gravity that it plays an important role in the quantum theory.\footnote{Outside LQG, a non-trivial role for $\g$ has been formally argued \cite{FreidelS} as the parameter that controls the width of torsion fluctuations, thus allowing to interpolate between a low-energy phase where the metric is the fundamental variable, and a high-energy phase where the connection degrees of freedom are fundamental. 
Mild evidence in this direction can be gathered from a 1-loop renormalization of $\g$ \cite{IoDario}.}

The action contains an internal tensor,
$P=(\Id +\g\star)/2$, that for $\g=\pm i$ is the projector on the left-handed (antiself-dual) representations of the Lorentz algebra.
As a consequence, choosing these special values one can formulate the theory in terms of chiral variables, a fact noted already by Plebanski \cite{Plebanski,Capo2}, and at the root of Ashtekar's Hamiltonian formulation \cite{Ashtekar:1987gu}.
It can be puzzling at first sight that a chiral theory can describe both helicities of the graviton, and be equivalent to the parity-even Einstein's theory. The answer to both questions lies in the key fact that the fundamental, chiral variable is complex. Suitable reality conditions are needed to recover general relativity, and it is the reality conditions that restore both helicities and a parity-even theory, see e.g. \cite{AshtekarBook}. While the reality conditions are well under control in the classical theory, they have so far provided difficult to quantize. The problem is sidestepped choosing $\g\in\R$, and giving up the self-duality of the fundamental variables. One can still describe the phase space in terms of SU(2) connections, but the SU(2) group is now to be seen merely as a auxiliary structure, non-trivially embedded as a manifold in the cotangent bundle of the Lorentz group. 
The real phase space variables thus obtained are related to the ADM ones by a canonical transformation parametrised by $\g$,\footnote{To be more precise, the canonical analysis of \Ref{S} shows the presence of secondary constraints, which are solved by a 2-parameter family of connections. One parameter can be fixed requiring commutativity of the connection, and the result is the $\g$-labelled family of so-called Ashtekar-Barbero connections. The secondary constraints also provide the way in which this connection can be (non-trivially) embedded in the space of Lorentzian connections. Notice that the family contains also a unique solution which has a more direct relation to a Lorentzian connection, but it is not commutative \cite{AlexandrovChoice}. }
and correspond to the same kinematical phase space of an SU(2) gauge theory. 

This phase space can be quantised in a way that makes no reference to a background spacetime metric, and the result is the Hilbert space of spin networks with its holonomy-flux algebra. The states are labelled by an oriented graph $\G$, and quantum numbers $j$ associated to links and nodes, and representing values of the gravitational field, that is values of the metric itself: for instance, areas, angles and volumes associated with loci dual to the graph.
The geometry so defined is quantum in three different senses: first, the geometric spectra are discrete, with minimal eigenvalue proportional to the Planck length; second, different geometric operators do not commute, making the geometry fuzzy; and third, for a given state on a given graph, only a finite number of degrees of freedom is available \cite{IoCarloGraph}.
In some sense, fixing a graph is similar to fixing the number of particles in Fock space, and 
the complete Hilbert space of the theory is recovered via a projective limit on a arbitrarily fine graph,
a procedure defining the background-independent Ashtekar-Lewandowski measure  \cite{ALmeasure}. 
It is often more convenient to think of the limit as a sum over all possible abstract graphs, and put to the side questions concerning non-trivial topologies and knot classes. This change of perspective is possible provided $(i)$ a state with zero labels on a link is equivalent to a state on a graph with that link removed (a property referred to as cylindrical consistency), and $(ii)$ dividing by the orbits of spatial diffeomorphisms does not leave behind any moduli (a more subtle property, that depends on the particular mathematical details of the initial embedded graph \cite{CarloWinston}).  

Two approaches are possible for studying the dynamics: canonically, the kinematical spin network states should be constrained by the action of the operator version of the Hamiltonian constraint, projecting in this way to a Hilbert space of physical states. That the Hamiltonian constraint is a well-defined operator is one of the major successes of LQG, and this program for the dynamics is being pursued in particular by Thiemann's group (see e.g. \cite{ThiemannAQG1}). 
In a more covariant framework, the projector on the physical states can be implemented via the spin foam formalism, that is a version of the functional integral used in QFT adapted to the discrete quantum structures at hand \cite{CarloZako,PerezLR}. The dynamics written in this way is a sum over histories of spin networks, each of them weighted by a quantum amplitude related in a precise limit to exponentials of the action \Ref{S}. That is, the spin foam formalism provides a constructive definition of the 
path integral appearing in the sum-over-geometries approach to quantum gravity,
\be
Z = \int {\cal D}g_{\m\n} \, e^{\f i\hbar S[g]} \ \mapsto \ Z = \sum_{\s} \sum_{j\in \s} {\cal A}_\s[j].
\ee
The currently most developed model goes by the acronym of EPRL \cite{EPRL}, and has a number of important features, some of which we briefly mention below.
A spin foam sum can be split into two pieces: first, a sum over all 2-complexes $\s$ (`foams') compatible with the spin network's graph on their boundary; second, a sum over the quantum numbers $j$ associated with the individual foam. 
Notice that the amplitude is strictly dimensionless: $\hbar G$ appears only when one is to match the result of the theoretical calculation with observations. Quantum amplitudes on a given foam can be explicitly `evaluated' as a saddle point expansion for large quantum numbers. This procedure unveils, for simplicial foams, the Regge action plus quantum corrections \cite{BarrettLorAsymp}.
The presence of the Regge action is promising for recovering the correct semiclassical limit, and the quantum corrections can be proved to be UV finite, and furthermore also IR finite if a cosmological constant is included \cite{Han}.

On the other hand, a key open question concerns the finiteness of the sum over foams. A mathematical tool to handle the summations is a special quantum field theory defined on a group manifold \cite{DePietri}. Such \emph{group field theories} reproduce spin foam amplitudes as their expansion in Feynman graphs. The last few years have seen a wealth of development in this area through the reinterpretation of group field theories as tensor models \cite{Oriti:2013aqa,Rivasseau:2013uca,GurauRyan11}, and the discovery of a 1/N expansion associated with it \cite{Gurau1/N10}. 
An alternative way to control the sum is to look at the refinement process and continuum limit, a line of thought pushed in particular by Dittrich's group (e.g. \cite{Dittrich:2012jq,Dittrich:2013xwa}, see also \cite{EteraCG13}).
A priori, one could focus on understanding the sum of foams first, and postponing a geometric understanding of the quantum theory to a later stage. However, the encouraging results mentioned above on the spin foam amplitudes at fixed foam have spurred some interest in understanding better the geometry of the theory on a fixed graph.

Fixing the graph corresponds to a truncation of the theory to a finite number of degrees of freedom, and as it turns out, the classical limit of a spin network on a fixed graph describes a notion of discrete geometry called twisted geometry \cite{twigeo}. A twisted geometry is a collection of Euclidean polyhedra, each of them dual to a node of the graph, plus an extra angle associated to each face. This information can be used to reconstruct a notion of intrinsic and extrinsic discrete 3-geometry, and the characteristic that makes it `twisted' is the fact that while each face as a well-defined area, its shape can differ when viewed from the adjacent polyhedra sharing it, so that the reconstructed metric is in general discontinuous. This lack of shape matching is what makes these geometries more general than Regge geometries.\footnote{Associating a discrete geometry to a spin network is not a unambiguous process, in some ways comparable to choosing an interpolating function given a finite data set \cite{IoCarloGraph}. The selecting criterium for twisted geometries is piecewise flatness. An alternative proposal, insisting on linearity and removing the discontinuities of twisted geometries, has appeared in \cite{FreidelSpinning}.} 
One can argue that in the continuum limit the shape mis-match becomes negligible, and then one recovers general relativity as in Regge calculus. This is really a dynamical question, so much depends on how precisely the Einstein's equations are to be implement on twisted geometries. The latter is an open line of research, whose application is particularly important for the understanding of spin foam models on arbitrary foams. A proposal for a generalised twisted Regge action has appeared in \cite{FreidelJeff13}. On the other hand, the shape matching conditions are automatically satisfied at the saddle point of the EPRL model, so it is enough to show that the Regge action dominates the behaviour to understand its semiclassical limit. 
This is a remarkable result, and provides strong interest in the model, and in the open questions lingering behind it.

The most pressing one concerns the evaluation and, if possible at all, geometric interpretation of the quantum corrections. Even if they are finite, they may spoil the continuum limit of the classical Regge action. This is a double problem of studying the asymptotic series provided by the saddle point expansion and the resummation/refined of foams. There are interesting questions also concerning the saddle point itself, and its relation to the Regge action. In particular, it has been pointed out that the saddle point equations are not compatible with curvature of the Regge equations, thus leading to a flatness problem \cite{HellmannFlatness}. Furthermore, the saddle point behaviour of the model on a non-simplicial foam has not being computed, and it is not even clear what it should look like.
These and other reasons motivate an effort to improve the geometric understanding of spin foam amplitudes, and twistors can be the right tool for this investigation, as I now describe.

\section{Twistors}
\label{sec-1}

The truncation of loop quantum gravity to a fixed, oriented graph, defines the Hilbert space ${\cal H}_\G$ of spin network states on $\G$, equipped with the holonomy-flux algebra, whose $\hbar\mapsto 0$ limit gives the phase space $S_\G := \times_l T^*\SU(2)/\!/C_n$, equipped with the canonical Poisson brackets of $T^*\SU(2)$, and where $/\!/$ means symplectic reduction by the Gauss constraint $C_n$ acting on the nodes of the graph.
The results of \cite{twigeo2,EteraTamboSpinor,EteraSU2UN,WielandTwistors,IoTwistorNet,IoWolfgang} show that twistors can be used as variables for $S_\G$, and that when written in terms of twistors, the geometric meaning of $S_\G$ becomes much more transparent, opening the way to progress on the dynamical and semiclassical understanding of the theory.

A twistor can be described as a pair of spinors,\footnote{The presence of an $i$ differs from the standard Penrose notation, and it is just a matter of convenience to bridge with the conventions used in loop quantum gravity. See \cite{PenroseRindler2} for an introduction to twistor theory.} 
$Z^\a=(\omega^A,i \bar\pi_{\dot A})\in\mathbb{C}^2\oplus\bar{\mathbb{C}}^2{}^\ast=:\mathbbm{T}$,
equipped with canonical Poisson brackets,
\be\label{piom}
\big\{\pi_A,\omega^B\big\}=\delta^B_A.
\ee
The space carries a representation of the Lorentz algebra, which preserves the complex bilinear
$\pi_A\om^A\equiv \po$,
under which $\omega^A$ is left-handed and  $\bar\pi_{\dot{A}}$ right-handed.
The space carries also a representation of the larger algebra $\su(2,2)$, the double cover of the conformal algebra of Minkowski,
which 
makes them useful to describe the degrees of freedom of massless particles, and that is the root of their recent blossoming in use in S-matrix theory. In particular, introducing the dual $\bar Z_\a= (-i\pi_A, \bar\om^{\dot A})$, 
the quantity $
s= Z^\a \bar Z_\a/2= \im\,\po
$
defines an $\su(2,2)$-invariant twistor norm that can be identified with the helicity of the particle. 

Twistors can be given a geometric interpretation via the incidence relation
\be
\om^A = i X^{A\dot A} \bar\pi_{\dot A},
\ee
that identifies a locus of points $X$ for each twistor.
In general, this locus describes a totally null plane in complexified Minkowski space.\footnote{An interpretation of this locus in real Minkowski space can also be provided in terms of the Robinson congruence of light rays, whose visual representation gives a clear meaning to the name twistor.} 
In the special case of null twistors, $s=0$, the locus is instead a light ray in real Minkowski space

Both algebraic and geometric properties enter crucially in the relation between twistors and loop quantum gravity. On the one hand, the algebraic properties can be generalised to carry a representation of the Lorentz holonomy-flux algebra, a feature necessary to capture the full phase space of LQG. On the other hand, the planes and null rays identified by twistors are used to reconstruct 
the twisted geometries.

Consider then an oriented graph $\G$, decorated with a pair of twistors on each link $l$. We denote $Z$ (`untilded') the twistor associated with the source, and $\tl Z$ (`tilded') the one associated with the target, and $\T_\G$ the resulting phase space. 
The key point to bridge between twistors and loop quantum gravity is that $S_\G$ is obtained from $\T_\G$ by symplectic reduction. This reduction hinges on three different types of constraints. While each of them is self-motivated by the resulting structure, they are also motivated by the discretised action principle used in spin foam models \cite{PerezLR}.

\begin{description}
\item[Area matching constraints:]
on each link, the two twistors are required to match their Lorentz-invariant parts,
\be\label{defC}
C := \po - \tpo =0.
\ee
Symplectic reduction by this constraint reduces $\T^2$ to $T^*\SL(2,C)$. This symplectic reduction, and the explicit parametrization of holonomies via twistors, was introduced and studied in \cite{twigeo2,WielandTwistors,IoTwistorNet,IoWolfgang}.
Notice that \Ref{defC} involves $\re(\po)$, a non-conformal invariant quantity. Imposing this constraint is responsible for reducing the initial conformal algebra of twistors to its $\SL(2,\C)$ sub-algebra.

\item[Simplicity constraints:]
for each twistor, we constrain the incidence relation by requiring
\be\label{inc}
X^{A\dot A} = r e^{i\th/2} N^{A\dot A}, \qquad r \in\R,
\ee
with $N^2=-1$. This fixed time-like vector should be thought of as a discretised version of the normal appearing in the 3+1 splitting of the gravitational action, whereas the extra phase is related to the Immirzi parameter via
$$e^{i\th} = \f{\g+i}{\g-i}.$$
The constraint \Ref{inc} corresponds to a discretised version of the (primary) simplicity constraints of the action \Ref{S}, and
has two immediate consequences. 
From an algebraic viewpoint, it restricts the generators of the Lorentz algebra to be simple as bi-vectors, up to the $\th$ phase.
That is, picking the canonical choice $N^I=(1,0,0,0)$, and denoting $\Pi^i=\Pi^i(\om^A,\pi^A)$ the left-handed generators of $\sl(2,\C)$ reconstructed from the twistor, \Ref{inc} implies 
\be\label{simpl}
\Pi^i = -e^{i \th}\bar\Pi^i.
\ee
Hence, left-handed and right-handed geometric structures are constrained to match, precisely as in the continuum theory.\footnote{See for instance \cite{MikeLR, Iobimetric} for the `left=right' interpretation of the simplicity constraints. The constraints \Ref{simpl} can also be rewritten in terms of boost and rotation generators as $K^i+\g L^i=0$, a form more familiar in the literature.}
From a geometric view point, it selects a real space-like plane out of the complex $\a$-plane.\footnote{This plane can be identified with the plane orthogonal to the spatial part of the null pole of $\om^A$.} 
The extra phase $\th$ plays an important role in the theory, and it vanishes only in the limit $\g\mapsto\infty$,
for which the canonical transformation from ADM to Ashtekar variables becomes singular.
Due to this phase, the twistor satisfying \Ref{inc} is not strictly speaking null. It is however trivially isomorphic to a null twistor, with isomorphism depending on $\g$, and we denote it $\g-$null.

\item[Closure constraints:]
on each node $n$ of the graph, we impose the closure condition 
\be\label{clos}
\sum_{l\in n} \om^A_l \pi_{lB} = \f{\sum_l (\po)_l}2 \, \d^A_B.
\ee
equivalent to say that $\sum_{l\in n} \Pi^i_l = 0$. In the latter form, it can be recognised as the discrete version of the Gauss law, as it appears for instance in lattice gauge theories.
For the purpose of the symplectic reduction on the total graph, it is important that the time-like vector $N^{I}$ is the same for each half-link sharing the same node. Then \Ref{inc} selects a unique space-like hypersurface around that node, and \Ref{clos} gives a Euclidean closure condition for the rotation generators, $\sum_{l\in n} L^i_l=0$. 
This equation defines a unique, bounded and convex flat polyhedron dual to the node, with number of faces $v_n$ equal the valence of the node, and the symplectic reduction gives a phase space of shapes of the polyhedron \cite{IoPoly}.

\end{description}

The resulting system of constraints on $\Gamma$ is fairly simple to study: all constraints are first class, except for the non-Lorentz-invariant components of \Ref{inc}, which form a second class pair on each half-link.
The reduced phase space has dimensions $2L+\sum_n 2({\rm v}_n-3) = 6L-6N$, and can be explicitly shown to be 
symplectomorphic to $S_\G$ \cite{IoWolfgang}.
Furthermore, the explicit reduction procedure reveals a convenient parametrization of the phase space in terms of a collection of polyhedra plus an extra angle associated to each shared face. This information describes the twisted geometries introduced earlier. 

The structure can be quantised proceeding \`a la Dirac, starting from an auxiliary Hilbert space carrying a unitary representation of the constraints, and implementing them as operator equations. The result is a equivalent representation for LQG, where instead of using projective spin networks as orthonormal basis, one uses homogeneous functions in spinor space.
Notice that the whole construction uses the full twistor space $\T$, and not the projective twistor space ${\mathbbm P}\T$ that has a prominent role in other applications of the theory. Here the scale of $\po$, which is the gauge removed when considering the projective space, describes the area of the face of the polyhedron, and upon quantization, reproduces the famous discrete spectrum of LQG.

The description in terms of twistors of the Lorentzian phase space of LQG allows us to reformulate also spin foam amplitudes as integral in twistor space, see \cite{IoWolfgang} and \cite{WielandHamSF}. Rewritten in these variables, the geometric meaning of the amplitudes is manifest, as well as their local Lorentz invariance, unitarity and complex structures. 
The twistorial description enriches the spin foam formalism of new tools which provide precious for both doing calculations and investigate the geometric and physical meaning of the path integral. 
For the sake of the reader familiar with recent results in the field, let me briefly mention a few open questions that I believe can be addressed in this formalism. First, the question of flatness.
While the saddle point Regge behaviour of the EPRL correctly captures the discretised Einstein's dynamics, 
some results (e.g. \cite{HellmannFlatness}) show that only the flat solutions are compatible with the saddle point approximation,
thus raising a question on the validity of the model, and on how to go beyond it. 
As shown in \cite{IoWolfgang}, the twistorial description allows to explicitly study the torsion tensor on a spin foam, and relate the flatness problem to the proper implementation of the torsion-free condition, since one can distinguish the Ashtekar-Barbero and the Levi-Civita holonomies. Secondly, the same improved geometric understanding can be applied to the case of non-simplicial foams and their saddle point approximation.

A more fundamental issue concerns the explicit evaluation of `radiative' corrections to the Regge action, and their behaviour under refinement of the foam.\footnote{If they spoil the semiclassical limit, the large spin limit involved in this analysis would bear no physical meaning, and the only way in which the formalism could make sense is via a different limit, to be suggested by the group field theory/tensor model resummation.}
Such corrections are extremely hard to evaluate, and very little is known about them (see e.g. \cite{Riello13}).
As it turns out, the EPRL amplitudes are not holomorphic functions of twistors. This suggest to seek for a formulation in terms of holomorphic functions instead, by means for instance of the Penrose transform, which could make the evaluation of the integrals determining the quantum corrections more tractable.
Finally, let me stress again that the connection between LQG and twistors is at present made graph by graph, namely order by order in an expansion on refined spin foams.
An important open question concerns the continuum limit of this description. 
In particular, the finite number of degrees of freedom and the piecewise-flatness play an important role in mapping LQG variables to twistors, and it is non-trivial to understand if such continuum limit would reproduce hypersurface twistors or not.
The work of Penrose and the more recent developments (e.g. \cite{AdamoMasonSkinner11}) show that twistors do capture the degrees of freedom of the continuum theory. If a continuum description of LQG in terms of twistors is also possible, this could provide an entirely new bridge between LQG and more conventional QFTs.

\section{The problem of time, the light-front formalism and some perspectives}
\label{sec-1}

As discussed earlier, there is a pre-problem of time, which is the identification of constraint-free data, and then the real task is to understand how the usual notion of dynamics is to be reconstructed out of such `frozen' Dirac observables.
Now, constraint-free data can in principle be found if one chooses to work with a null hypersurface instead of a space-like one.
The free data can be the metric on a 2-dimensional foliation of an outgoing asymptotic light cone \cite{Bondi62, Sachs62, FrittelliNull95}, or part of the Weyl tensor \cite{NP62}, or part of the Ashtekar variables \cite{Goldberg92}.
Notice also that in such a formulation, the Hamiltonian constraint becomes second class: In fact, it can not generate gauge transformations as there are no infinitesimal deformations of a null hypersurface that keep it null.
While these are important result, the situation is far from simple.
Using null hypersurfaces data has a long history in classical and quantum field theory,
and many open questions have slowed down the promising potential associated with it  (see e.g. \cite{LesHouchesNull}).
Specific problems to the null formulation of general relativity involve expressing the kinematical variables in terms of the free data, that is reconstructing the 4-dimensional Lorentzian metric, controlling the development of caustics, the possible presence of zero-modes, etc. Research is active (albeit not very popular) in this domain, and it is not my intention here to review it, but rather to point out a specific direction where this formalism, together with the twistorial description of LQG, can bring something new to the problem of time.

Specifically, once the constraint-free data are available, one can attempt a Schr\"odinger-like quantization on the reduced phase space \cite{AshtekarBook,CarloNull96}. However, it is far from clear how to properly implement quantization rules on a phase space that contains the time evolution within itself, and the program has not received much attention. 
On the other hand, the spin foam formalism is more amenable to a covariant treatment, and appears suitable to attack this problem, particularly using the twistorial description described above.
Indeed, we have seen how twistors elegantly represent the (primary) simplicity constraints as the constrained incidence relation \Ref{inc}. Written in this form, there appears to be nothing special about a time-like $N^I$, and one can equally consider a null vector. This null case, and the associated symplectic reduction, were studied in \cite{IoNull}, and the results can be summarised as follows.
The Euclidean polyhedra of twisted geometries are replaced by null polyhedra with space-like faces, and SU(2) by the little group ISO(2). The main difference is that the simplicity constraints present in the formalism are all first class, 
with the new gauge symmetries corresponding to the freedom of moving the polyhedra along the null direction without changing their intrinsic geometry. As a consequence, information on the shapes of the polyhedra is lost, and the result is a much simpler, abelian structure given by the helicity subgroup of the little group. At the gauge-invariant level, such null twisted geometries can be described by an Euclidean singular structure on the 2-dimensional space-like surface defined by a foliation of space-time by null hypersurfaces. The quantization of the phase space gives a notion of spin networks for null hypersurfaces. Such spin networks are labelled by SO(2) quantum numbers, and are embedded non-trivially in the unitary, infinite-dimensional irreducible representations of the Lorentz group. 

These results clarify the kinematical structure of null boundary states for spin foams.
The next step for research in this direction is to construct a spin foam model that carries such null data on its boundary, and to investigate its semiclassical properties and quantum dynamics. This is also motivated by the many applications such a spin foam model can have, starting with black hole physics.
Twistors will prove extremely useful to investigate this line of research, and allow us to see whether the spin foam formalism can shed new light over the quantum version of the problem of time.

\subsection*{Acknowledgements}
I would like to thank the organisers of the conference for the invitation, and 
Sergey Alexandrov and Alejandro Perez for discussions on the light-front formalism.



\end{document}